\documentclass[12pt,preprint2]{aastex}

\usepackage{epsfig}

\shorttitle{Reconsidering the origin of the 21\,{}micron feature}
\shortauthors{Posch et al.}

\begin{document}

\title{Reconsidering the origin of the 21\,{}micron feature:
 Oxides in carbon-rich PPNe? \footnote{Based 
on observations with ISO, an ESA project with instruments funded 
by ESA Member States (especially the PI countries: France, Germany, 
the Netherlands and the United Kingdom) and with the participation 
of ISAS and NASA.}}

\author{Th. Posch}
\affil{Institut f\"ur Astronomie, T\"urkenschanzstra{\ss}e 17, A-1180 Wien, Austria}
\email{posch@astro.univie.ac.at}

\author{H. Mutschke}
\affil{Astrophysikalisches Institut,
Schillerg\"asschen 2-3, D-07745 Jena, Germany}

\and

\author{A. Andersen}
\affil{NORDITA, Blegdamsvej 17, DK-2100 Copenhagen, Denmark}


\begin{abstract}

The origin of the so-called `21'\,$\mu$m feature which is especially prominent
in the spectra of some carbon-rich protoplanetary nebulae (\mbox{PPNe}) is 
the matter of a lively debate. A large number of potential band carriers
have been presented and discarded within the past decade. The present
paper gives an overview of the problems related to the hitherto 
proposed feature identifications, including the recently suggested 
candidate carrier silicon carbide. We also discuss the case for 
spectroscopically promising oxides.

SiC is shown to produce a strong resonance band at 20--21\,$\mu$m 
if coated by a layer of silicon dioxide. At low temperatures, core-mantle 
particles composed of SiC and amorphous SiO$_2$ indeed have their strongest 
spectral signature at a position of 20.1\,$\mu$m, which coincides with the 
position of the `21'\,$\mu$m emission band.

The optical constants of another candidate carrier that has been relatively 
neglected so far -- iron monoxide -- are proven to permit a fairly accurate 
reproduction of the `21'\,$\mu$m feature profile as well, especially when 
low-temperature measurements of the infrared properties of FeO are taken 
into account. As candidate carrier of the `21'\,$\mu$m emission band,
FeO has the advantage of being stable against further oxidation and
reduction only in a narrow range of chemical and physical conditions, 
coinciding with the fact that the feature, too, is detected in a small group 
of objects only. However, it is unclear how FeO should form or survive
particularly in carbon-rich \mbox{PPNe}.

\end{abstract}



\section{Introduction}

\citet{KVH89} reported the discovery of a strong emission band,
centered (as it then seemed) at 21\,$\mu$m, in the spectra of four carbon-rich
protoplanetary nebuale (\mbox{PPNe}) taken by the Infrared Astronomical Satellite
(IRAS). The feature was not resolved well with the Low Resolution Spectrometer 
(LRS), but nevertheless, it was evident already from the LRS observations that it
has a considerable strength in some objects (especially IRAS 07134+1005)
and is too broad (full width at half maximum $>$2\,$\mu$m) to originate 
from any atomic transition. The pioneering work by Kwok, Volk \& Hrivnak 
also highlighted the fact that the `21'\,$\mu$m feature only occurs in a 
short transitional phase of stellar evolution, i.e.\ in a very limited range of 
physical and chemical conditions. Any identification of the `21'\,$\mu$m 
feature has to cope with this fact.

Observations of selected \mbox{PPNe} with the Infrared Space Observatory (ISO)
by \citet{HVK00} brought substantial progress, especially concerning the 
position and width of the feature, which was shown to be 20.1\,$\mu$m
instead of 21\,$\mu$m (notwithstandig, the feature is still being refered to as
`21'\,$\mu$m band, a tradition to which we also stick in this paper).
Hrivnak, Volk \& Kwok (2000) {analysed} ISO observations at 2-45\,$\mu$m for 
seven \mbox{PPNe} and two other carbon-rich objects. Six of the sources exhibit 
emission features at 21 and 30\,$\mu$m. The authors resolved the 30\,$\mu$m band 
into a broad feature peaking at 27.2\,$\mu$m (the `30'\,$\mu$m feature) and a 
narrower feature at 25.5\,$\mu$m (the `26'\,$\mu$m feature). The observations 
suggest that the sources with `21'\,$\mu$m feature do also show the 30\,$\mu$m
feature. The `26'\,$\mu$m feature appears to be present in a broader range
of objects -- carbon stars, \mbox{PPNe}, and possibly PNe -- than does the 
`21'\,$\mu$m feature.

The candidate carriers of the `21'\,$\mu$m band have since its discovery 
been the matter of a long discussion.
Cox (1990) presented IRAS-LRS-spectra of ten H\,II regions which showed a 
peak close to 21\,$\mu$m. He associated this peak with the feature discovered
by Kwok, Volk \& Hrivnak and tentatively assigned it to iron oxides with the
stoichiometries $\gamma$-Fe$_2$O$_3$ and Fe$_3$O$_4$. In the late 1990s, some 
of these H\,II regions have also been observed with ISO; in the ISO 
spectra, however, we do not find any traces of a broad 21 or 
20\,$\mu$m emission band (in accordance with Oudmaijer 
\& de Winter (1995), who already cast doubt on Cox' observations).

\citet{Goe93}, still on the basis of IRAS spectra, assigned the `21'\,$\mu$m 
feature to silicon disulfide (SiS$_2$), which has indeed an intense vibrational
band in the 20\,$\mu$m spectral range (but, unfortunately, a second one
which is not observed; see below, Sect.\ \ref{disc}). From a chemical point 
of view, other sulfur compounds like S$_8$, OCS, S$_2$ and CS$_2$, even 
though mainly as mantles of other grains, were also taken into account 
by Goebel as potential emitters in the 20\,$\mu$m region.

{\citet{Web95} brought fulleranes as potential carriers of the 
`21' \,$\mu$m feature into the discussion.}
\citet{Hill98}, having studied the IR bands of terrestrial nitrogen-doped 
and neutron-irradiated diamonds, {showed}\/ that defects in diamond 
structures can cause an emission band at 20$-$22\,$\mu$m and even 
stronger bands in the 8--10\,$\mu$m region. However, there is a 
significant discrepancy between the position of 21--22\,$\mu$m 
and the actual position of the `21'\,$\mu$m band as 
determined by recent observations.

\citet{Pap00} calculated synthetic spectra of carbonaceous macromolecules
containing OH, oxygen, sulfur, nitrogen or a combination of them and found
that it is possible to reproduce not only the `21'\,$\mu$m feature, but also the
3.3, 11.3, 12+, 26 and 30\,$\mu$m bands by emissions of such
molecules. According to this scenario, the `21'\,$\mu$m feature would 
arise from out-of-plane vibrations of oxygen or nitrogen atoms in 
5-membered carbon rings. Papoular's approach seems promising
because it {aims at}\/ an identification of {\em all}\/ the major features 
detected in the sources of the `21'\,$\mu$m band on the basis of {\em one}\/ 
unified dust model. But at the time being, substantial problems
remain, such as the unsatisfactory reproduction of the observed 
26--35\,$\mu$m band profile.

There are various other papers discussing carbonaceous materials
as carriers of the `21'\,$\mu$m feature, but without any detailed fits of
its average profile. {\citet{Sour92} suggested amides, especially urea, 
as potential carrier, \citet{Buss90} and \citet{Just96} suggested polycyclic 
aromatic hydrocarbon molecules and \citet{Gris01} hydrogenated amorphous 
carbon.}

Von Helden et al.\ (2000) came up with another suggestion,  
attributing the 20.1\,$\mu$m feature to titanium carbide clusters 
(made of 27 to 125 atoms). Spectra of bulk titanium carbide and TiC particles
as those subsequently published by \citet{HM01} and \citet{KK03} show no 
band in the respective wavelength region, but rather weak bands at 9.5, 12.5 
(Kimura \& Kaito 2003) and 19\,$\mu$m (Henning \& Mutschke 2001; 
see also Speck \& Hofmeister 2004). 
This does not contradict the measurements by \citet{Helden2000}, 
however, since clusters are likely to have electronic and vibrational 
properties deviating from those of bulk material even if composition 
and crystal structure are similar.

Notwithstanding, in 2003, three papers doubting the `TiC-cluster-identification'
with substantial arguments came out, all based essentially on arguments
concerning the abundance of titanium (Hony et al.\ 2003, Chigai et al.\
2003, and Li 2003). \citet{Hony2003} argue that TiC, like
any kind of dust, can only emit as much in the infrared as it absorbs in
the UV and visual range. Assigning a simple model opacity to TiC at short
wavelengths, Hony et al.\ found that this energy balance can be fulfilled 
only if unrealistically large amounts of TiC are predicted to form
in carbon-rich \mbox{PPNe}. 
This point was strengthened by \citet{Li03} by applying a sum-rule 
based on Kramers-Kronig considerations on the wavelength-integrated 
absorptivity. \citet{Cetal03} came to the same result, comparing the
number densities of SiC and TiC grains necessary for producing the 
11\,$\mu$m and 21\,$\mu$m features, respectively, with the available Si 
and Ti abundances.

Recently, \citet{SH04} reported the experimental
finding that silicon carbide, under certain circumstances, does not only show
the well-known resonance feature at 11\,$\mu$m, but also a secondary band 
(not predicted by group theory), which is centered at 20--21\,$\mu$m.
This secondary band is reported to appear only in the $\beta$-SiC-polytype, 
and nitrogen or carbon impurities were suspected by the authors to favour its occurence. 
Cold, maybe N- or C-doped $\beta$-SiC-dust (with temperatures around 100\,K) 
would then, according to Speck \& Hofmeister, emit at 20--21\,$\mu$m (depending on 
particle size and shape) even more effectively than at 11\,$\mu$m, since the 
latter peak would still be in the Wien regime of the corresponding Planck
function.
{Important as this suggestion might be for the identification of the 
`21'\,$\mu$m band, it is not beyond any doubt, as we try to show in \S 3.}

The present paper is structured as follows. 
In \S 2, we summarize the currently available astronomical information
on the sources in which the enigmatic `21'\,$\mu$m band is observed.
\S 3 presents new insights to the infrared properties of materials 
which should, despite some problems, {\em not}\/ be excluded as feature 
carriers -- such as core-mantle particles composed of SiC and SiO$_2$ 
and cold FeO dust grains. In \S 4, we deliver additional arguments for 
ruling out potential carriers of this band which have been ruled out 
more or less convincingly by other authors. The last section (\S 5) 
summarizes our conclusions.


\section{Properties of the sources of the `21'\,$\mu$m feature 
\label{s:prop}}

So far, the `21'\,$\mu$m feature is observed only in emission, only in 
objects with a C/O-ratio {close to or greater than} unity and only in 
{\em cool}\/ dusty environments, i.e.\ environments with dust temperatures below 250\,K.
Among them are the already mentioned C-rich \mbox{PPNe} (about half of them show the
feature), two Planetary Nebulae \citep{HWT01} {with central stars of the Wolf-Rayet 
type} and, according to \citet{VXK00}, two extreme carbon stars (IRAS 21318+5631, 
IRAS 06582+1507). \citet{Clem04}, however, re-analyzing the ISO-spectrum of 
IRAS 21318+5631, find no evidence for a feature at 20\,$\mu$m, and in the case 
of IRAS 06582+1507, \citet{VXK00} also remark that its reality is doubtful.

It should be noted that an emission band with a position and width quite comparable
to the `21'\,$\mu$m band is detected in the spectra of stars on the {\em early}\/
Asymptotic Giant Branch (AGB), i.e.\ at the very opposite end of AGB evolution:
the 19.5\,$\mu$m feature \citep{P02}.
Its peak position amounts to 19.4--19.6\,$\mu$m, {its profile is symmetric}\/
and its full width at half maximum (FWHM) is close to 3\,$\mu$m. The dust shells 
of the mostly semiregular variable stars in which this feature is observed are 
generally assumed to consist of refractory oxides, but not at all of carbonaceous
dust grains. {A solid solution of FeO and MgO has been proposed as feature 
carrier in this case (see also \citet{FG03}).}

\begin{figure}[htbp]
\epsscale{.80}
\plotone{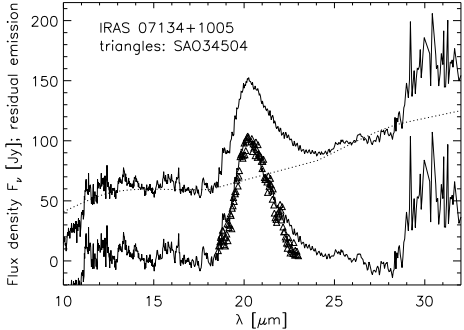}
\caption{ISO-SWS spectrum of the PPN IRAS 07134+1005
at a resolution $\lambda$/$\Delta\,\lambda$ = 600. At the bottom, 
the profile of the `21'\,$\mu$m band is shown, both as derived from 
the ISO-SWS spectrum HD 56126 (solid line) and from an 
ISO-SWS spectrum of SAO 34504 (open triangles) by subtraction of 
a polynomial (dotted line overplotted to the ISO spectrum at the top).
\label{ISO1}}
\end{figure}

Coming back to the \mbox{PPNe} exhibiting the 21\,$\mu$m band in its `canonical' 
form, let us consider the spectrum and the derived properties of its best-known 
source, IRAS 07134+1005 (=SAO 96709 or HD 56126, see Fig.\ \ref{ISO1}). 
The F5 I central star of this PPN has a surface temperature of about 
7000\,K according to \citet{HVK00}; other \mbox{PPNe} sources of the same feature have
comparable effective temperatures (5500 to 8000\,K). The circumstellar envelopes of these
objects are all {\em optically thin}\/ at infrared wavelengths. As for unusual
properties of the central stars, \citet{VW00} found them to be enriched 
in s-process elements {and metal-poor}. Apart from the peak due 
to the radiation of the central star at visual wavelengths, several infrared 
bands arising from dust grains are present in the spectrum of HD 56126, namely 
at 8, 11.3, 12.1, and 30\,$\mu$m. They are partly unidentified, partly 
attributed to SiC (11.3--broad component), PAHs (11.3--narrow component, 12+) 
and magnesium sulfide (30\,$\mu$m), respectively. No silicate features 
are seen in any of the sources of the `21'\,$\mu$m band.

We derived the profile of the `21'\,$\mu$m band using the polynomial subtraction 
method which has been applied by \citet{P99} for the derivation of the 
13\,$\mu$m band profile. The residual emission of HD 56126, based on an
ISO-SWS-spectrum reduced with the ISO Spectral Analysis Package (ISAP), 
is shown at the bottom of Fig.\ \ref{ISO1}. For comparison, the scaled
residual emission of SAO 34504 -- calculated by the same method -- is shown 
in the bottom part of Fig.\ \ref{ISO1} as well (open triangles). The close
coincidence of the two emission profiles is noteworthy; {together with
the result by \citet{VKH99} that the feature profile is pretty much the 
same in all sources, this implies that the spectrum of HD 56126 can be
considered representative and will be so considered in the following.
The most conspicuous properties of the feature profile are a steep rise 
in its blue wing, a maximum position of 20.1\,$\mu$m and a long 
`tail' extending to 23--24\,$\mu$m.}

In contrast to stars on the early AGB, bright \mbox{PPNe} have spatial extensions 
which are large enough to resolve them even at infrared wavelengths. \citet{KVH02}
and \citet{Miy03} tried to answer the question whether the different broad 
infrared features of \mbox{PPNe} have their origin in different or coincident 
regions on the basis of imaging through broad filters. They came to the 
conclusion that the 11.3\,$\mu$m SiC, the 12+\,$\mu$m plateau emission, and 
the `21'\,$\mu$m emission originate (approximately) in the same region, which 
seems to indicate that similar physical and chemical conditions are required 
for the formation and/or survival of the respective feature carrier.


\section{Silicon carbide, silicon dioxide and iron monoxide \label{s:oxides}}

\subsection{Silicon carbide with impurities or coatings}

{As mentioned in \S 1, \citet{SH04} suggested silicon carbide with
impurities as carrier both of the 11.3 and the `21'\,$\mu$m spectral
features. One of their arguments in favour of this hypothesis is the
spatial co-location of both emission bands in some of the objects exhibiting
them. Obviously, this argument is applicable not only to N-doped and C-doped 
SiC, but also to any other particular SiC grain population like e.g.\ oxidized
SiC grains (see below).}

In the following, when using the term `silicon carbide' or `SiC',
we will always refer to the cubic (beta) modification of this polymorphous
solid. Due to its symmetry, $\beta$-SiC has only one predicted infrared active
mode, which is located at 796\,cm$^{-1}$ (position
of the maximum of $\epsilon''$), producing, in the case of small
spherical particles, a maximum of its absorption efficiency factor
at 935\,cm$^{-1}$ or 10.7\,$\mu$m. (For elongated ellipsoidal particles,
this maximum of the absorption efficiency is shifted towards longer wavelengths: 
see Mutschke et al.\ 1999 and below.) As already explained, secondary 
emission bands of SiC can occur in the case of a non-ideal crystal 
structure, e.g.\ due to carbon or nitrogen inclusions. 
{The case for N-doped SiC, for which Speck \& Hofmeister refer to \citet{Sutt92},
is problematic, however, since Suttrop et al.\ did not only detect a 20.9\,$\mu$m 
(478\,cm$^{-1}$) band, but also stronger additional bands in N-doped SiC.
These additional bands (e.g.\ at 667\,cm$^{-1}$ (15\,$\mu$m) and 580\,cm$^{-1}$ 
(17.2\,$\mu$m)) are not seen in the sources of the astronomical `21'\,$\mu$m feature. 
Moreover, the whole series of nitrogen-induced secondary features of SiC 
is weaker than the SiC main feature at least by a factor of 1000. Therefore,
very low temperatures (below 80\,K) would be necessary to reverse this
ratio of band strenghts.
According to \cite{HVK00}, the dust temperatures at the {\em inner}\/ radii of
the dust shells of the `21'\,$\mu$m sources are between 165 and 220\,K;
due to the relatively small ratio between outer and inner shell radius,
the {\em average}\/ temperature of the dust grains producing the `21'\,$\mu$m
will hardly be below 80\,K even for very transparent dust species
(see below, Fig.\ \ref{TR-det}).

As for C-doped SiC, there is pre\-sent\-ly too little information at hand
about its potential secondary bands. As long as no studies on analytically
well-characterized SiC samples with carbon impurities are available, it 
remains unclear whether or not the properties of N-doped SiC can be 
extrapolated to C-doped SiC.}

\begin{figure}[htbp]
\plotone{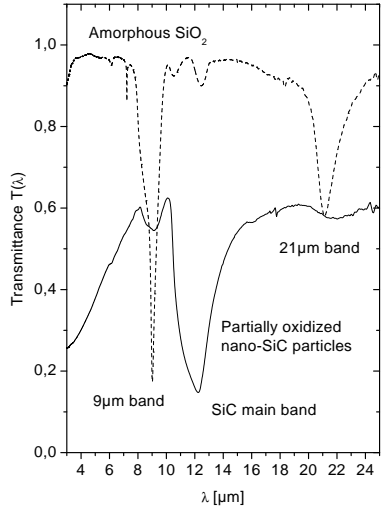}
\caption{Comparison of the transmission spectrum of partially oxi\-dized 
nano-SiC particles with that of amorphous SiO$_2$ grains.
 \label{Lab-SiC}}
\end{figure}

\begin{figure}
\plotone{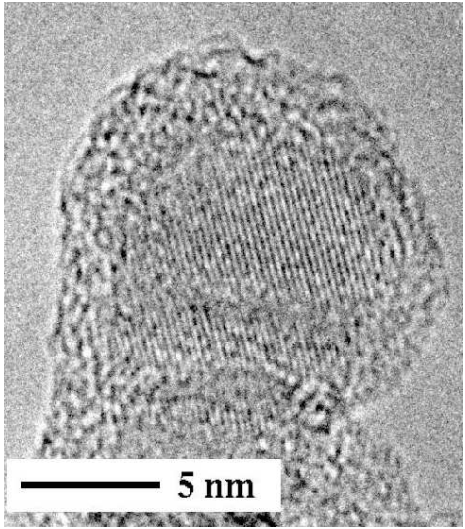}
\caption{Core-mantle particle consisting of SiC and amorphous
silicon oxide. The silicon carbide core is identified by its characteristic
lattice fringes.  \label{TEM}}
\end{figure}

Interestingly, there is yet another possibility to produce an 11 plus
a `21'\,$\mu$m band by, so to speak, `SiC plus something': namely by
coating SiC grains with SiO$_x$.

\begin{figure}[htbp]
\plotone{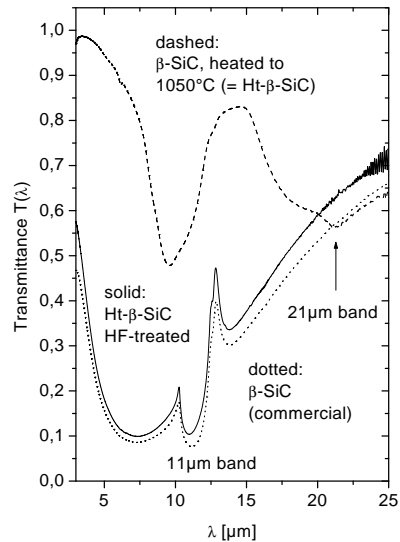}
\caption{Commercial $\beta$-SiC (dotted line) was heated in air
at 1050$^{\circ}$\,C (dashed) and has afterwards been treated with HF, thereby
returning from an almost completely oxidized state (carrying a 9\,$\mu$m
and a strong 21--22\,$\mu$m absorption band) to its original state characterized 
by the well-known 11\,$\mu$m band.
The additional broad 7\,$\mu$m and 14\,$\mu$m absorptions in SiC are due to 
plasmon-phonon coupling, see \citet{M99} for details.
 \label{HF-SiC}}
\end{figure}

\begin{figure}[htbp]
\plotone{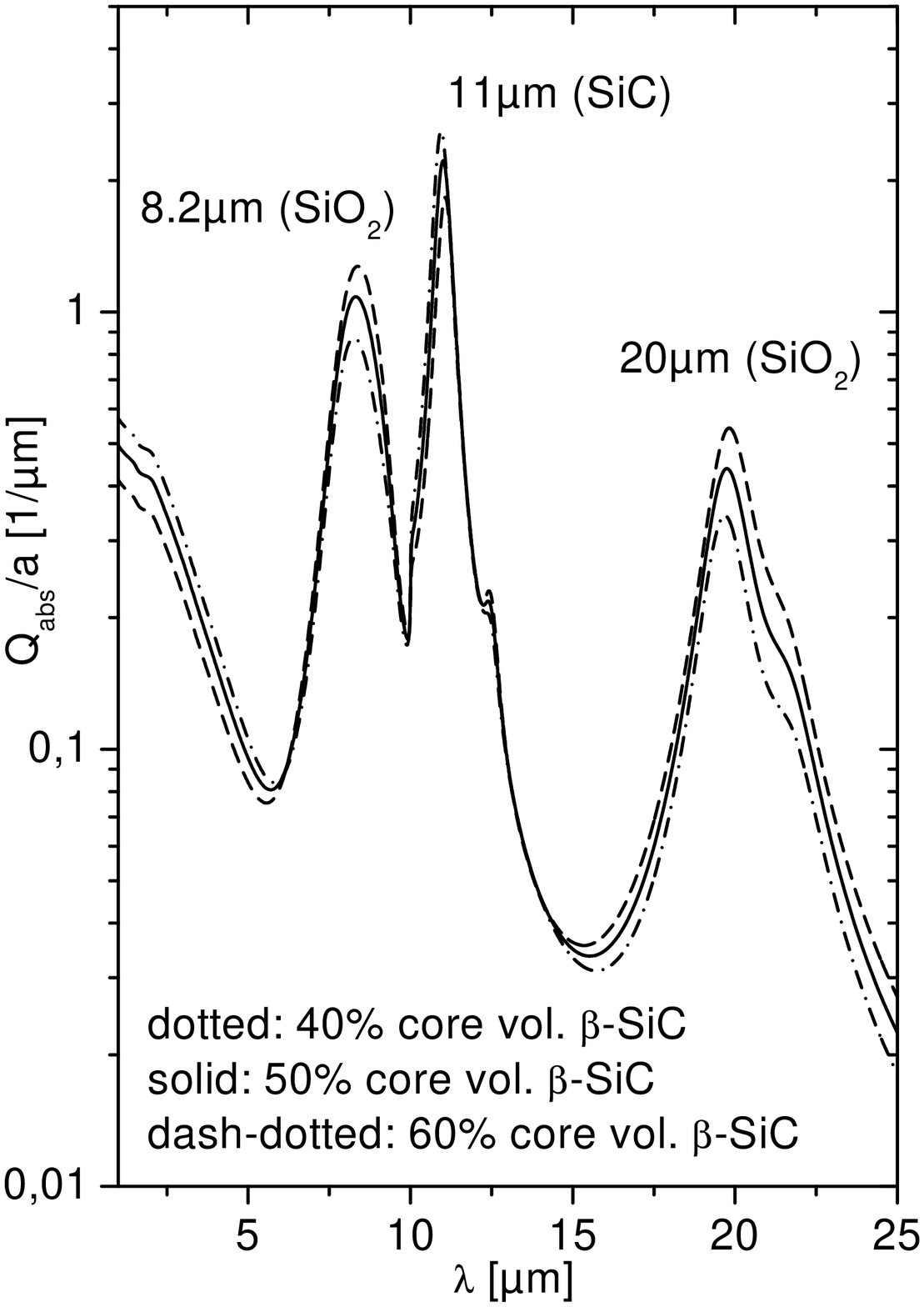}
\caption{Volume normalized absorption efficiency Q$_{\rm abs}$/a of core-mantle
particles composed of an SiC core and an amorphous SiO$_2$ mantle,
the size of the whole (spherical) grains being small compared to the 
wavelength.
\label{SiCSiO}}
\end{figure}

Recently performed experiments {\citep{Clem03}} de\-mon\-strate that 
pure SiC nanoparticles, under the influence of the laboratory atmosphere, 
are quite easily oxidized at their surfaces. This effect is shown in 
Fig.\ \ref{Lab-SiC}, where we compare the transmittance of SiO$_2$ grains 
with the transmittance of partially oxi\-dized SiC nanoparticles (both embedded 
in KBr). The method for the production of SiC nanoparticles by laser pyrolysis 
and the infrared spectra of these grains have been discussed extensively 
by \citet{Clem03}. They also report the detection of an amorphous silicon 
oxide mantle by means of transmission electron microscopy (see Fig.\ \ref{TEM}). 
The large surface-to-volume ratio of the SiC core favours the formation and 
growth of a silicon oxide mantle, which can be a few nanometers thick and 
thus reach a considerable volume fraction of the particulate. Further evidence 
for the formation of SiO$_2$ on SiC surfaces, has been delivered --
even though for significantly higher temperatures -- by \citet{Mend02} 
and \citet{Tang03}. 

Depending on the particle size and shape as well as agglomeration effects, 
SiC nanoparticles show one strong IR resonance band, located at 10.8--12.4\,$\mu$m. 
However, if partially oxi\-dized at their surfaces, the SiC nanograins produce additional 
infrared bands at 9 and 21--22\,$\mu$m. These bands can be assigned to stretching 
and bending vibrations of the SiO$_4$ tetrahedra. The minor 12.5\,$\mu$m band of 
SiO$_2$ is not detected in the spectrum of partially oxi\-dized SiC nanoparticles, 
since it almost coincides with the SiC main band shifted to longer wavelengths 
by shape effects.

An additional proof for the oxidation of SiC is delivered by treating
the oxidized grains with HF, which quickly reduces the oxidized grains,
leaving only the pure silicon carbide cores. The spectroscopic result of 
this process is depicted in Fig.\ \ref{HF-SiC}.

Not only experimentally, but also by calculating the absorption efficiency
of core-mantle-particles composed of SiC and amorphous SiO$_2$, absorption
spectra with the principal structure of those shown in Figs.\ \ref{Lab-SiC}
and \ref{HF-SiC} can be derived. For this purpose, the optical constants 
of amorphous SiO$_2$ compiled in the first volume of \citet{Palik85}
have been used. We performed a Lorentz fit to these optical
constants in order to analyse the underlying oscillator parameters.
The damping constants of the SiC oscillator has been set
to 50\,cm$^{-1}$ and that of the 456\,cm$^{-1}$
oscillator of SiO$_2$ to 42 (instead of 27), assuming that the nanocrystalline
structure is comparable to that of defect-bearing polycrystalline films 
(see \citet{M99}, Sect.\ 2.3). As for the 456\,cm$^{-1}$
SiO$_2$ oscillator, a larger damping constant than that corresponding
to the Palik data is justifyable also on the basis of the spectra published
by \citet{LSH84}. These authors studied the transmittance of thin SiO$_{\rm x}$
films (with x$<$2) and found a strong broadening of the 20--21\,$\mu$m
band with decreasing x. This result is of potential relevance for the coating 
of SiC by silicon oxide in carbon-rich circumstellar shells, since in these 
environments, the formation of oxygen-deficient SiO$_2$ is very well
conceivable. For a summary of all Lorentz parameters used, see Tab.\ \ref{Lor1}.

\begin{table}[htb] 
\caption{Frequencies ($\rm TO_j$), oscillator strengths ($\rm \Omega_j$)
and damping constants ($\rm \gamma_j$) of SiC and SiO$_2$.
We adopted the values
$\rm \epsilon_{\infty}\,=\,6.49$ for SiC and
$\rm \epsilon_{\infty}\,=\,2.2$ for amorphous SiO$_2$.
\label{Lor1}}
\begin{center} 
\begin{tabular}{l|l|l|l} 
\hline        
\centering
                & $\rm TO_j$  & $\rm \Omega_j$ & $\rm \gamma_j$ \\ 
                & (cm$^{-1}$) & (cm$^{-1}$)    & (cm$^{-1}$)    \\ 
                \hline
  SiC           & 796 & 1423 & 50     \\ 
  				\hline 
  SiO$_2$	    & 456  & 417  & 42    \\
                & 798  & 150  & 30    \\
                & 1080 & 866  & 60    \\
                & 1194 & 257  & 100   \\    
                & 1479 & 200  & 80    \\
                \hline  
\end{tabular} 
\end{center} 
\end{table}

Fig.\ \ref{SiCSiO} shows the absorption efficiency factor Q$_{\rm abs}$/a of 
spherical core-mantle-particles composed of SiC and amorphous SiO$_2$ with core
volume fractions ranging from 40\% to 60\%.
Three prominent maxima of Q$_{\rm abs}$ are discernible: a peak
at 8.3\,$\mu$m, which originates from the SiO$_2$ mantle; the sharp 11\,$\mu$m 
SiC resonance; and, finally, a 20\,$\mu$m peak, with a main peak and a 
shoulder at the long-wavelength side, originating again from the SiO$_2$ mantle. 
Peak splitting is typical for bands of mantle materials. It arises from the existence 
of two interfaces with core and ambient medium.
All the bands are sharper than those of the transmission spectrum of
partially oxi\-dized SiC shown in Fig.\ \ref{Lab-SiC}. This difference in the
bandwidth is primarily due to shape effects.

\begin{figure}[htbp]
\plotone{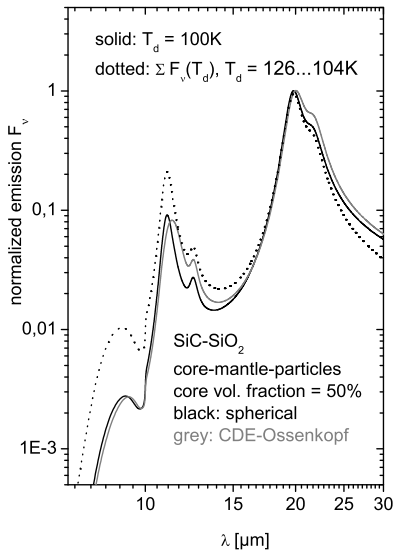}
\caption{Products of the absorption efficiency for SiC-SiO$_2$ core-mantle particles 
(core volume fraction = 0.5) with two different Planck functions, corresponding
to dust temperatures of 160\,K and 100\,K. In grey colour,
the case of ellipsoidal particles (shape distribution according to Ossenkopf,
Henning \& Mutschke 1992) is overplotted. \label{PlSiCSiO}}
\end{figure}

Looking at Fig.\ \ref{SiCSiO}, one may assume that the depicted pattern
of C$_{\rm abs}$/V maxima is incompatible with the spectra of astronomical objects
like HD 56126, since the 20\,$\mu$m band, in this figure, is only a minor peak
compared to those at 8.2 and 11\,$\mu$m. However, such a conclusion would be
premature given that the dust in the sources of the `21'\,$\mu$m feature is
rather cold, as has been highlighted in Sect.\ \ref{s:prop}. For dust particles
with temperatures below 150\,K, all features at wavelenghts smaller than
$\sim$\,20\,$\mu$m are strongly attenuated, because they tend to be located
in the Wien regime of the corresponding Planck functions. This effect, which was
already noted by \citet{SH04}, is illustrated in Fig.\ \ref{PlSiCSiO}.

\begin{figure}[htbp]
\plotone{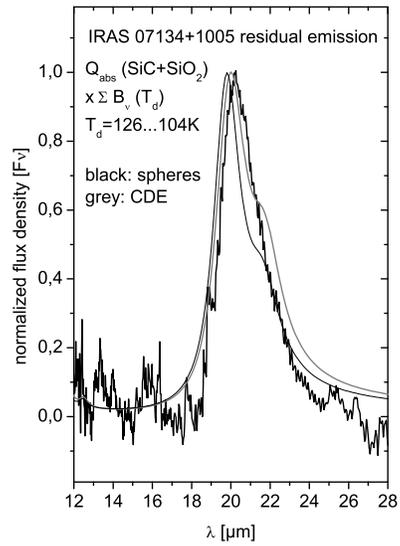}
\caption{{Normalized emission spectrum of a dust shell composed of 
SiC-SiO$_2$ core-mantle grains. The shell is supposed to follow a
1/r$^{2}$ density distribution and a temperature distribution resulting
from the condition of radiative equilibrium (see appendix) and a certain
geometry of the shell of HD 56126 as infered from observational data 
(see below, sect.\ 3.3). The residual emission of HD 56126 is shown 
for comparison.}
\label{SiCvs20my}}
\end{figure}

It is evident from this figure that cold core-mantle-particles composed
of SiC and amorphous SiO$_2$ do produce a strong 20\,$\mu$m feature.
There is a rather good agreement between the observed and the calculated 
band positions {(see also Tab.\ \ref{t:delta-tab} for a quantitative
assessment)}, and the weakness of the observed 11\,$\mu$m feature 
can be explained by dust temperatures significantly lower than 160\,K on the 
average. There remains a problem, however, with the {\em width}\/ and 
{\em shape}\/ of the 20\,$\mu$m band produced by SiO$_2$. For spherical 
particles, the FWHM of the 20\,$\mu$m SiO$_2$ emissivity maximum amounts 
to 1.9\,$\mu$m, compared to an observed FWHM of 2.2--2.3\,$\mu$m. 
Furthermore, the shoulder at the `red' wing of the 20\,$\mu$m SiO$_2$ 
feature has not been seen in the astronomical `21'\,$\mu$m band. The 
bandwidth discrepancy can be overcome by a CDE; the shoulder at the `red' 
wing of the band, however, does not vanish in the CDE case (see Fig.\ 
\ref {SiCvs20my}); {nor does it vanish by any particular (realistic)
choice of the dust temperature or dust temperature distribution.}


\subsection{Oxidized silicon grains}

Instead of growing on top of silicon carbide grains, SiO$_{2}$ may equally well
form mantles on top of pure silicon particles. \citet{LD02} were the first to 
exa\-mine the spectra emerging from silicon grains with oxide coatings, focussing 
on the case of stochastically heated nanoparticles. {\citet{SW02} proposed
silicon nanoparticles as carriers of the so-called {\em extended red emission}\/
observed in circumstellar environments; however, this identification is
still a matter of debate.} 

The principal structure of the IR spectrum of Si-SiO$_2$ core-mantle-grains is
similar to that of SiC-SiO$_2$ core-mantle-grains, except for the absence of
the 11\,$\mu$m band. In addition to the 8.2\,$\mu$m and the 20\,$\mu$m 
peaks, a rather strong 12.5\,$\mu$m band appears in the Si-SiO$_2$ 
absorption efficiency spectrum. The exact position $\lambda_{20}$
of the 20\,$\mu$m resonance depends on the volume fraction $f$ of the 
Si core. $\lambda_{20}$ increases from 19.7 to 19.8 \,$\mu$m as
$f$ decreases from 0.6 to 0.4.

\begin{figure}[htbp]
\plotone{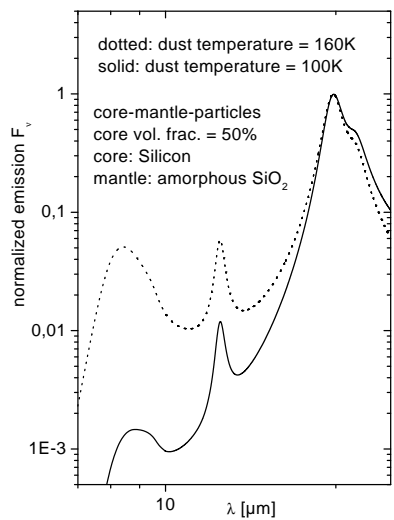}
\caption{Products of the volume normalized absorption efficiency C$_{\rm abs}$/V 
shown in the previous figure with two different Planck functions, corresponding
to dust temperatures of 160\,K and 100\,K. 
\label{PlSiSiO}}
\end{figure}

Following the procedure that has been explained in the previous subsection,
we folded the absorption efficiency of a Si-SiO$_2$ core-mantle-grain
with $f$\,=\,0.5 with two different Planck curves B$_{\nu}$(T$_{\rm d}$). 
For the dust temperature, the two values T$_{\rm d}$\,=\,{100}\,K 
and T$_{\rm d}$\,=\,{160}\,K have been chosen. As for the 20\,$\mu$m band, 
the result -- plotted in Fig.\ \ref{PlSiSiO} -- strongly resembles the 
case of SiC cores, which is not surprising, since the optical constants 
of the core have only a limited influence on the structure of those bands 
which are due primarily to the mantle. 

It must be stressed, however, that core-mantle particles composed of
Si and SiO$_2$ are more unlikely to be the carrier of the `21'\,$\mu$m
band than those composed of SiC and SiO$_2$, since it is questionable whether 
{\em silicon}\/ grains can be formed in carbon-rich \mbox{PPNe} at all.


\subsection{Cold FeO (wustite)}

As mentioned in \S 1, iron oxides have been considered as carriers of
bands in the 20--21\,$\mu$m region by various authors already. It seems
clear, though, that none of the `higher' iron oxides like Fe$_2$O$_3$
or Fe$_3$O$_4$ can survive in a reducing environment. Iron monoxide
(FeO, also called wustite\footnote{Strictly speaking, wustite is the
terrestrial form of iron monoxide and has the chemical composition 
Fe$_{\rm 1-x}$O with x\,=\,0.05 \dots 0.16.}), by contrast, may survive 
in a carbon-rich circumstellar shell. 

According to \citet{Dul80}, the oxidation of metallic iron is a very
efficient process provided that enough O$_2$ molecules are present.
{The degree to which iron particles are oxidized depends on the physical 
conditions, inter alia on the temperature. The {\em higher iron oxides}\/
Fe$_3$O$_4$ or Fe$_2$O$_3$ become stable at higher temperatures
(see, e.g., Gail \& Sedlmayr 1998). By contrast, as
\citet{WR61} and \citet{FM70} pointed out, several monolayers 
of {\em iron monoxide}\/ form very quickly on top of Fe surfaces, 
even at low temperatures, since this process requires a negligible activation 
energy. In order to end up with dust grains composed essentially of FeO 
(instead of a large metallic iron core and a tiny FeO mantle), the (low-temperature-)oxidation
of {\em very small}\/ iron grains is required -- i.e.\ grains composed of
10$^3$ atoms at maximum. Only such small iron grains -- their diameter
amounts to $\sim$2\,nm -- could be fully converted into FeO by
the process mentioned above. The oxidation of iron grains consisting
of less than 10$^3$ atoms is to some extent an artificial assumption;
less artificial, though, than the constraint that the TiC clusters --
which have previously been considered as carriers of the `21'\,$\mu$m
band -- should consist of no more than 125 atoms to efficiently 
produce the TiC cluster signature.}

Why would FeO, according to our suggestion, not exist in the vast majority of the 
PNe, at least not with high number densities? A plausible explanation is that in 
PNe with their large fractional ionization and high UV photon densities, 
iron oxides will be reduced to metallic iron. While the non-detection of the 
`21'\,$\mu$m band in PNe spectra can thus be accounted for, its non-detection 
in the spectra of most of the oxygen-rich as well as carbon-rich AGB stars 
is more difficult to understand within the iron grain oxidation scenario for 
PPNe. If in PPNe like HD 56126, FeO forms by oxidation of iron, why does 
the same process not take place in C-stars, S-stars and in the vast 
majority of M-AGB stars? This may be due to the fact that the dust around these
stars is, on the average, much hotter than dust in PPNe and that the sticking 
probability of oxygen on iron decreases as the temperature increases. The 
Mg-Fe-oxide formation around some M-stars, postulated by \citet{P02}, must
involve an entirely different mechanism than the low-temperature oxidation
of iron in PPNe. Indeed, \citet{FG03} predict Mg-Fe-oxides via the condensation
of the very refractory MgO in M-stars. 
{It is noteworthy that HD 56126 and other sources of the `21'\,$\mu$m band 
studied by \citet{VW00} are depleted in metals including iron. 
Other PPNe do not show the feature, even though more iron is available for 
FeO formation. There could be two reasons for that. One is that the iron particles 
should be small as stated above. The second is that FeO formation is essentially 
governed by oxygen availability and that the iron abundance is therefore only 
crucial if there is enough free oxygen.}

\begin{figure}[htbp]
\plotone{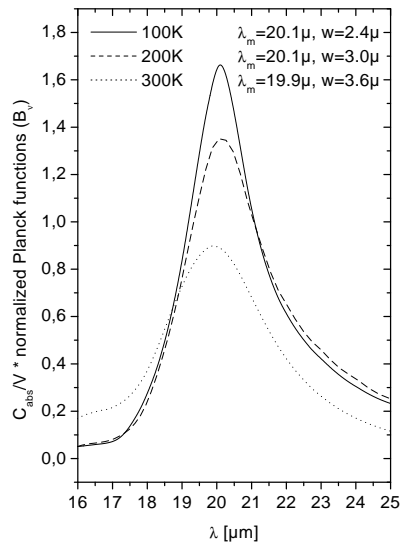}
\caption{Absorption efficiency factors C$_{\rm abs}$ of spherical
FeO grains, calculated from their optical constants measured at
room temperature (T\,=\,300\,K) and at low temperatures
(T\,=\,200\,K, T\,=\,100\,K). The respective C$_{\rm abs}$ functions
have been multiplied with Planck functions B$_{\nu}$ (T) 
(normalized at 20\,$\mu$m) for the appropriate temperatures.
\label{FeOTemp}}
\end{figure}

Whereas the knowledge on the Fe-Mg-oxide formation in circumstellar environments
is still scarce, quite detailed {\em spectroscopic}\/ information on Mg-Fe-oxides is available.
The optical constants of Fe$_{\rm x}$Mg$_{\rm 1-x}$O samples (x amounting 
to 0.4 \dots 1.0) with magnesiowustite structure have been measured at 
room temperature by \citet{H95}. Subsequently, \citet{HM97} published 
{\em low temperature data}\/ for various cosmic dust analogs, i.e.\ 
their indices of refraction n($\lambda$) and absorption k($\lambda$) 
for T\,=\,10, 100, 200 and 300\,K. A major change of the infrared spectrum 
of Fe$_{\rm x}$Mg$_{\rm 1-x}$O is observed as the samples are cooled: 
namely a sharpening of the vibration band located at 18--20\,$\mu$m 
(depending on $x$). Fig.\ \ref{FeOTemp} shows this effect 
for $x$\,=\,1.0, i.e.\ for pure FeO. (The case T\,=\,10\,K 
has been omitted, since it does sot significantly differ from 
T\,=\,100\,K.)

The FWHM of the 20\,$\mu$m FeO band decreases
from 3.6\,$\mu$m at room temperature to 2.4\,$\mu$m at T\,=\,100\,K,
while the band position is shifted from $\lambda_{\rm max}$\,=\,
19.9\,$\mu$m to $\lambda_{\rm max}$\,=\,20.1\,$\mu$m.
The latter effect is exclusively due to the multiplication with the
Planck functions B$_{\nu}$ (T), which have maxima positions
of 17, 25.5 and 51\,$\mu$m for T\,=\,300, 200 and 100\,K,
respectively. The rather large width of the 20\,$\mu$m band produced
by FeO, in combination with the fact that for T\,=\,200\,K 
and T\,=\,100\,K, the peak position is located at the `blue' wing of the 
corresponding Planck curve, leads to an asymmetric profile of the 
C$_{\rm abs}\times$B$_{\nu}$(T) curves -- the same kind of 
asymmetry which is found for the astronomical `21'\,$\mu$m band.

Iron monoxide has a rather high absorption efficiency in the visual range
of the spectrum. Therefore, it is comparatively strongly heated in the
radiation field of an F or G star. An equilibrium temperature of less than
100\,K is reached only at distances from the central star larger than 
2\,$\times$\,10$^{4}$ stellar radii. This can be shown by calculating
the energy balance for the case of an equilibrium between the radiation 
field of the central star and the dust grains in its circumstellar
{according to the scheme summarized in the appendix.}

According to \citet{KVH02}, the inner radius R$_{\rm in}$ of the dust shell of IRAS
07134+1005 is located at 1.2'' from the central star, corresponding to
1.2 (1.8) $\times$ 10$^{-2}$\, pc for an assumed distance of 2 (3) kpc
(\citet{KVH02} assume a distance of 2\,kpc, \citet{Meix03} assume 3\,kpc).
In terms of the stellar radius corresponding to a luminosity of 1400\,L$_{\odot}$ 
kpc$^{-2}$ \citep{HVK00} and a temperture of 7000\,K, this amounts to 
9760 or 6490 stellar radii for d=2 (3) kpc, respectively.

\begin{figure}[htbp]
\plotone{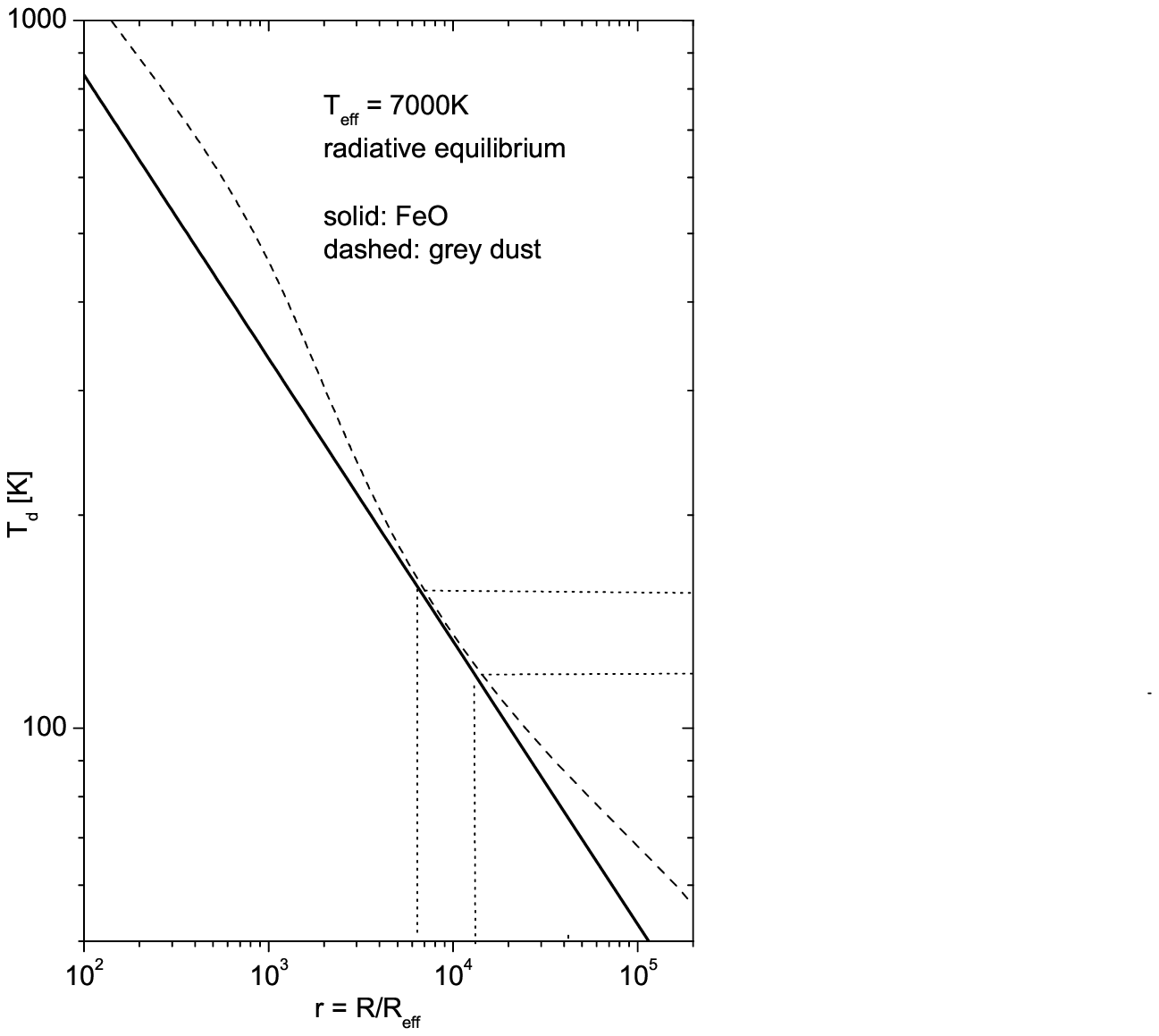}
\caption{Radiative equilibrium temperature of FeO in the radiation
field of a star with an effective temperature of 7000\,K (solid line).
For comparison, the radial temperature profile dust with wavelength-independent 
optical constants is also shown (dashed line; see also the appendix to this 
paper). Short-dotted lines delimit the r-range of the brightest part of IRAS 
07134+1005.
\label{FeOTR}}
\end{figure}

In Fig.\ \ref{FeOTR}, the position of  R$_{\rm in}$/R$_{\rm eff}$\,=\,6490
has been highlighted by a dotted vertical line. The equilibrium temperature of
FeO corresponding to this distance from the central star amounts to 155\,K.
This is in good agreement with the results of the model calculations by
\citet{HVK00}, who derive an inner temperature of the dust shell of 165\,K
by radiative transfer calculations aimed at a fit of the spectral energy distribution.

\begin{figure}[htbp]
\plotone{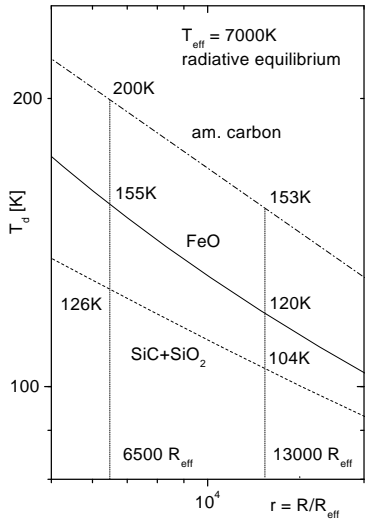}
\caption{Equilibrium temperatures of amorphous carbon, FeO and core-mantle
particles composed of SiC and SiO$_2$ in the radiation field of an F5 star. 
The domain from which presumably most of the IR emission of 
IRAS 07134+1005 comes is indicated by dotted lines.
\label{TR-det}}
\end{figure}

In terms of the ratio between outer and inner shell radius, the dust shells of 
the \mbox{PPNe} showing the `21'\,$\mu$m feature are rather thin.
For IRAS 07134+1005, the observations of \citet{KVH02} and \citet{Miy03}
yield an outer shell radius not much larger than twice the inner shell radius.
This implies that the dust temperature range, for a r$^{-0.4}$-like decrease
of T(r),\footnote{{See the appendix for a comment on the meaning of the
T(r) $\propto$ r$^{-0.4}$ decrease.}} is rather small. For FeO, 
we derive a dust temperature at 2 R$_{\rm in}$
of 120\,K, meaning that 120\,K$<$T$_{\rm d}$$<$155\,K (again for d=3\,kpc). 
For other dust species which are generally assumed to be present in carbon-rich \mbox{PPNe}, 
narrow T$_{\rm d}$ ranges are found as well. Amorphous carbon is warmer than FeO 
due to the monotonous increase of its absorption efficiency factor from the IR 
to the UV: we find 153\,K$<$T$_{\rm d}$$<$200\,K under the same conditions 
as above; core-mantle particles composed of SiC
and SiO$_2$, unless heavily N-doped, remain cooler than FeO 
(104\,K$<$T$_{\rm d}$$<$126\,K;\footnote{These values refer 
to a core volume fraction of 0.5; the larger the (rather transparent) 
SiO$_2$ mantle, the lower the dust temperature will be.} see Fig.\ 
\ref{TR-det}).

\begin{figure}[htbp]
\plotone{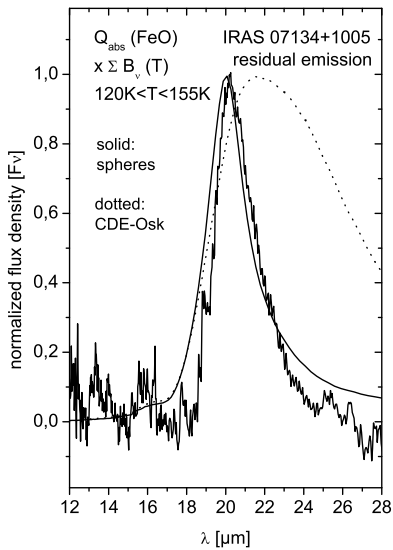}
\caption{Normalized emission spectrum of an FeO dust shell (solid: spherical particles,
dotted: CDE according to Ossenkopf et al.\ (1992)) with a temperature
distribution corresponding to the range highlighted in Fig.\ \ref{TR-det} and
a 1/r$^{2}$ density distribution. For comparison, the residual emission
of IRAS 07134+1005 is shown.
\label{FeOvsPPN}}
\end{figure}

These estimates make it possible to calculate the emission spectrum of an optically
thin dust shell with a homogeneous density distribution and a temperature range of
120\,K$<$T${\rm _d}$$<$155\,K. If we are only interested in the {\em profile}\/ 
of the emerging spectral energy distribution, it suffices to calculate a mean of blackbody 
functions B$_{\nu}$(T) for 120\,K$<$T$_{\rm d}$$<$155\,K and to multiply the result 
with C$_{\rm abs}$(T$_{\rm m}$).
The resulting normalized spectrum is shown in Fig.\ \ref{FeOvsPPN}. Both with respect 
to feature position and width, the emission spectrum of FeO calculated for the temperature
range indicated above reproduces the observed `21'\,$\mu$m band profile fairly accurately,
even though no {\em ad hoc}\/ assumptions concerning the dust temperature have been
made.

For a CDE, quite different spectra of FeO emerge. For a CDE according to 
\citet{Osk92} (`CDE-Osk'; it is characterized by a maximum probability for
particle shapes close to the spherical shape), the position of the FeO absorption
efficiency maximum is shifted to 20.7--21\,$\mu$m, and the bandwidth changes
even more dramatically, i.e.\ it increases to 6.5--7.6\,$\mu$m {(see Fig.\
\ref{FeOvsPPN}).}
For a CDE with equal probability of all particle shapes, the changes
in feature position and shape are even more dramatic, {but this kind
of shape distribution is unlikely to represent the shapes of
circumstellar grains. The assumption that nearly spherically symmetric
grains dominate in circumstellar dust shells is more reasonable, hence
the `truth' should lie between the `CDE-Osk' case and the case of spherical
particles.}

Fig.\ \ref{FeOvsPPN} may give the impression that FeO dust can produce a feature
comparable to the astronomical `21'\,$\mu$m emission band only under {\em very}\/
particular physical conditions, e.g.\ only for T$<$155\,K, while we know from
\citet{HVK00} that the dust temperatures in the sources of the `21'\,$\mu$m band
can reach 220\,K. However, as long as T$<$250\,K, B$_{\nu}$(T) will always have
its maximum at a wavelength larger than 20\,$\mu$m, thus shifting the 19.9\,$\mu$m
emissivity maximum of FeO to a feature position of 20.0--20.1\,$\mu$m. And indeed,
T$<$250\,K seems to be fulfilled for all known sources of the `21'\,$\mu$m feature.
In a wider sense, however, it is true that FeO can produce a strong 20.1\,$\mu$m
feature only under particular physical conditions. If the mean dust temperature is much
smaller than 100\,K, too little energy will be contained in the 20\,$\mu$m FeO band
to produce a feature as strong as that observed in some \mbox{PPNe}. If, 
by contrast, T$>$250\,K, the feature will become too broad and to much shifted 
to the `blue' as to account for the astronomical `21'\,$\mu$m 
band.\footnote{\citet{HM97} point out that only a limited number of 
cosmic dust analogs have optical constants which change as strongly at low
temperatures as those of FeO (and FeS).}

\subsection{Remark on the elemental abundances required to account for the `21'\,$\mu$m band}

Li (2003) made a very useful estimate of the mass of dust which is required to
account for the radiative energy contained in the `21'\,$\mu$m band in the case of
IRAS 07134+1005.
He pointed out that the Kramers-Kronig equation relates the integral over the 
absorption efficiency of a grain to its volume and the orientational averaged 
polarizability F. Since for all materials with a static dielectric constant 
$>$4 and for moderately elongated grain shapes, the latter quantity has
a value between 0.5 and 1 (Purcell 1969, Fig.\ 1), there is an upper 
limit to the integrated absorption efficiency per unit volume, which is more or less 
the same for the dust species considered here.
Consequently, there is also a general upper limit to the energy emitted in 
the 21\,$\mu$m band per unit particle mass or, vice versa, a minimum total grain volume 
for a certain total energy emitted through this band. Hence, the minimum mass 
derived by Li can be adopted to other materials just by scaling with the mass 
density of the grain material. 
This leads to the following conclusion concerning the abundance problem:
While there is, according to Li (2003), about 100 times too little amount of 
titanium to account for the '21'\,$\mu$m band by assigning it to TiC, there is 
2.5 times more Fe than would be needed to account for the '21' micron band 
with FeO (Fe is 250 times more abundant than Ti in HD 56126).
Sulfur is 312 times more abundant than Ti in the same star, silicon 1225 
times, carbon 34500 times and oxygen 36000 times.
Hence, compounds of iron, sulfur, silicon, carbon and oxygen cannot be
excluded as carriers of the `21'\,$\mu$m band in HD 56126 and objects
with comparable elemental abundances.


\section{Discarded feature carriers \label{disc}}

For the sake of completeness, two potential carriers of the `21'\,$\mu$m band 
which have been considered as promising candidates {(but should no longer be 
considered)} will be discussed: silicon disulfide and titanium carbide.
In the case of TiC, insurmountable problems with the abundance of titanium exist, 
whereas SiS$_2$ can be ruled out by spectroscopic 
arguments.


\subsection{Silicon disulfide}

As mentioned in \S 1, the formation of sulfides is very likely in carbon-rich
circumstellar environments. Indeed, the 26\,$\mu$m feature, observed in many 
sources of the `21'\,$\mu$m feature, is currently ascribed to magnesium 
sulfide (MgS, Hony et al.\ 2003). 

\begin{figure}[htbp]
\plotone{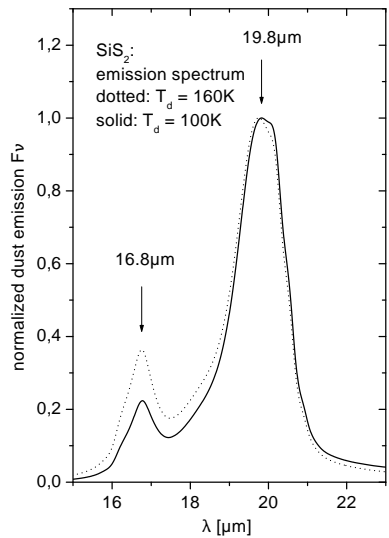}
\caption{Normalized emission spectrum of spherical SiS$_2$ particles
for two different dust temperatures, 100 and 160\,K. The 16.8\,$\mu$m
secondary emission band, which is not observed in the sources of the
`21'\,$\mu$m band.
\label{SiS2}}
\end{figure}

However, from a spectroscopic point of view, there are problems with ascribing
the 20.1\,$\mu$m band to SiS$_2$. The main point is that this material has
{\em two}\/ infrared bands, of which the second, located at 16.8\,$\mu$m
for spherical particles (see Fig.\ \ref{SiS2}), has no counterpart in 
the astronomical spectra. Furthermore, the position and FWHM of the main
SiS$_2$ band, which amout to 19.8 and 1.6\,$\mu$m, respectively, do not agree
with the observed characteristics of the 20.1\,$\mu$m band. For a CDE,
this discrepancy becomes smaller ($\lambda_{\rm max}$\,=\,20.2\,$\mu$m,
FWHM\,=\,1.8\,$\mu$m), but the problem with the unobserved secondary
feature at 16.8\,$\mu$m becomes even worse, given that the ratio between
$\sim$17 and $\sim$20\,$\mu$m band increases for a CDE.
Very low dust temperatures (significantly below 100\,K) would be required
to make the $\sim$17\,$\mu$m band negligible in strength compared to
the $\sim$20\,$\mu$m SiS$_2$ band. 

The visual and near-infrared optical constants of SiS$_2$ have not yet been derived,
such that it is not possible to calculate T(r) for SiS$_2$ based on its radiative
energy balance. However, it is hardly conceivable that SiS$_2$ is so transparent
in the visual range as to remain much colder than 100\,K over the whole range 
indicated in Fig.\ \ref{TR-det}.

\subsection{Titanium carbide}

Even though recent papers by \citet{Li03}, \citet{Cetal03} exclude TiC
nanoparticles as the carrier of the `21'\,$\mu$m band, we append a discussion of
TiC's infrared properties here, focussing on the relation between bulk TiC and
TiC clusters.

Macroscopic TiC has metallic properties, with an absorption index $k$
which is larger than the refractive index $n$ throughout the whole infrared range.
For $\lambda$$>$13\,$\mu$m, both $n$ and $k$ are larger than 10.
At about 19\,$\mu$m, a resonance peak occurs, which is rather weak,
however, with a relative increase of $n$, $k$ and C$_{abs}$ by less than 
10\%.

Small TiC {\em clusters}\/, e.g.\ those composed of 4$\times$4$\times$4
atoms, behave differently from bulk matter insofar as the mobility of the cluster
electrons is reduced. The lattice resonance is then no longer a small additional
term in the dielectric function, but the dominant contribution at infrared frequencies.
Therefore, a comparatively strong 20\,$\mu$m emission band arises from the
clusters; its exact position, however, depends on the cluster size.
\citet{Cetal03} derived the following Lorentz oscillator parameters for
the small TiC clusters produced by \citet{Helden2000}: a resonance frequency 
of 488\,cm$^{-1}$, a damping constant of 50.2\,cm$^{-1}$ and an 
oscillator strength of 187\,cm$^{-1}$. For spherical particles, this translates 
into a maximum position of the absorption efficiency factor of 20.0\,$\mu$m 
and an FWHM of 2.0\,$\mu$m. For a CDE, no shift of the band position is found
in this case (due to the non-negative values of the dielectric function of TiC 
according to \citet{Cetal03}), but only an increase of the FWHM to 2.2\,$\mu$m 
(for a CDE according to Ossenkopf, Henning \& Mathis 1992).

\begin{figure}[htbp]
\plotone{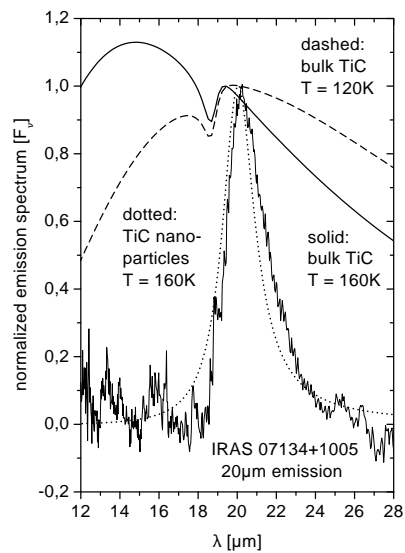}
\caption{Normalized emission spectrum of bulk TiC at temperatures
of 120\,K (dashed) and 160\,K (solid) and of TiC nanoparticles
at T = 160\,K (dotted) compared to the residual emission of 
IRAS 07134+1005.
\label{TiCvs20}}
\end{figure}

Fig.\ \ref{TiCvs20} shows the normalized emission of a small TiC 
cluster (at T${\rm_d}$=160K), the normalized dust emission arising 
from macroscopic dust grains (at two different temperatures T$_{\rm d}$ =160K 
and T$_{\rm d}$=120K) and, again for comparison, the profile 
of the astronomical `21'\,$\mu$m feature as derived from the ISO-spectrum 
of IRAS 07134+1005.
Since the maximum of the absorption efficiency factor of TiC is not evident
from Fig.\ \ref{TiCvs20} due to the normalization, it should be added that
C$_{\rm abs}$/V (TiC) does not exceed 0.6\,$\mu$m$^{-1}$. (Depending 
on the value of $\epsilon_{\infty}$, C$_{\rm abs}$/V may also be 
substantially smaller). By contrast, the $\sim$20\,$\mu$m peak 
absorption efficiency of cold (100\,K) FeO as well as that of amorphous 
SiO$_{2}$ amounts to 1.6\,$\mu$m$^{-1}$.

\section{Conclusions}

{Tables \ref{t:bandpos} and \ref{t:delta-tab} summarize the spectroscopic information
on the potential carriers of the `21'\,$\mu$m feature in two different respects.
Tab.\ \ref{t:bandpos} gives an overview of the positions of the absorption efficiency
maxima and of the corresponding bandwidths.
Tab.\ \ref{t:delta-tab}, on the other hand, refers to normalized dust emission spectra like
those shown in Figs.\ \ref{SiCvs20my}, \ref{FeOvsPPN}, \ref{SiS2} and \ref{TiCvs20},
hence to products of the absorption efficiency with blackbody functions. These products
are compared with the repeatedly shown profile of the `21'\,$\mu$m band: we present
the sums (henceforth: $\Sigma$$\delta^{2}$) of the squares of the differences between the `profile' 
and the `dust emission' values, for wavelengths between 17 and 25\,$\mu$m (outside this range, 
the ISO profile is characterized by comparatively large noise). The `temperature grid' has 
been chosen such as to cover the range between 160\,K and 70\,K in steps of 10\,K.
The bold numbers indicate the respective minima of $\Sigma$$\delta^{2}$.
For SiC+SiO$_2$-sph, TiC-sph and FeO-sph, the minimum of $\Sigma$$\delta^{2}$ 
occurs at blackbody temperatures between 80 and 90\,K. This is due to the
asymmetric shape of the profile of the `21'\,$\mu$m band. On the basis of a
symmetric aborption efficiency profile, this asymmetric profile is best reproduced
by multiplication with a blackbody of 80--90\,K. If, however, the absorption
efficiency of the feature carrier has an {\em intrinsically}\/ asymmetric profile
(as for ellipsoidal SiC+SiO$_2$-core-mantle grains), it is not necessary to assume
such a low dust temperature (120\,K are sufficient in this case).
The $\Sigma$$\delta^{2}$ values for SiS$_2$ are particular insofar as they continuously
decrease with decreasing temperature, down to less than 70\,K. This has a very 
simple reason: The 16.8\,$\mu$m secondary band of SiS$_2$ (see Fig.\ \ref{SiS2}), 
which has no counterpart in the astronomical spectra, is the more suppressed 
the lower the temperature. As regards the absolute values of $\Sigma$$\delta^{2}$, 
they are smallest for TiC clusters and second smallest for ellipsoidal SiC+SiO$_2$
core-mantle-grains.}

\begin{table*}[htbp]
\begin{center}
\caption{Overview of the most promising carriers of the `21'\,$\mu$m feature
which have been discussed in the present paper. Note: The observed position
and FWHM of the `21'\,$\mu$m band are 20.1\,$\mu$m and 2.3\,$\mu$m, respectively.
`sph' means spherical particle shape, `CDE-Osk' designates a continuous distribution
of ellipsoids according to \citet{Osk92}. The $\lambda$ (C$_{\rm abs, max}$) values
do {\em not}\/ directly represent feature positions, since these result from the product
C$_{\rm abs}$$\times$B$_{\nu}$\,(T$_{\rm d}$); for dust temperatures close
to 150\,K, this leads to a shift towards the `red' of 0.1--0.2\,$\mu$m.
\label{t:bandpos}}
\begin{tabular}{lccc}
\hline\hline
candidate carrier    & particle shape & $\lambda$ (C$_{\rm abs, max}$) [$\mu$m] & FWHM (20) [$\mu$m] \\
\tableline                                   
SiC+SiO$_2$		     & sph.$^{(a)}$    & 8.1--8.3, 11.0--11.1, 12.5, 19.7--19.8  & 1.9  \\
				     & CDE-Osk$^{(a)}$ & 8.2--8.4, 11.1--11.2, 19.8--20.0   & 2.1--2.6  \\
Si+SiO$_2$$^{(a)}$   & sph.		       & 8.0, 12.5, 19.7--19.8  & 1.8   \\                                
FeO$^{(b)}$          & sph.			   & 19.9             & 2.3         \\
                     & {CDE-Osk}       & {20.7}	          & {6.5}  \\
SiS$_2$   	         & sph.            & 16.8, 19.8       & 1.6  \\
 					 & CDE-Osk         & 16.8, 20.2       & 1.8  \\
TiC clusters		 & sph             & 20.0             & 2.0  \\
 					 & CDE-Osk         & 20.0      		  & 2.2  \\
TiC grains           & sph			   & 19.2             & --$^{(c)}$ \\
\tableline
\end{tabular}
\tablenotetext{(a)} {core-mantle-particles with SiC core volume fractions between 0.4 and 0.6}
\tablenotetext{(b)} {at an average temperature of $\sim$ 140\,K}
\tablenotetext{(c)} {broad bump, FWHM not defined}
\end{center}
\end{table*}

\begin{table*}[htbp]
\begin{center}
\caption{Selected potential carriers of the `21'\,$\mu$m feature: quality of fits
to the profile of the `21'\,$\mu$m band for different {\em mean} dust temperatures T$_d$
(between 130\,K and 70\,K). As an indicator for the quality of an individual fit,
the sum of the squares of the differences `normalized 21\,$\mu$m flux'
minus `C$_{\rm abs}$ $\times$ B$_{\nu}$\,(T$_{\rm d}$)' has been calculated
for 17\,$\mu$m $<$ $\lambda$ $<$ 25\,$\mu$m. This sum corresponds to the usual
`$\chi^{2}$' insofar as the quantities subtracted from each other are already
normalized.
\label{t:delta-tab}}
\begin{tabular}{lccccc}
\hline\hline
T$_d$ [K] & SiC+SiO$_2$-sph$^{(a)}$  & SiC+SiO$_2$-CDE$^{(a)}$  & FeO-sph & TiC-sph & SiS$_2$-sph \\  
\tableline                                   
160   & 5.98         & 3.10       & 6.40       & 3.54       & 18.3 \\
150   & 5.66         & 2.92       & 6.03       & 3.29       & 17.8 \\
140   & 5.31         & 2.80       & 5.62       & 3.02       & 17.4 \\
130   & 4.94         & 2.71       & 5.20       & 2.72       & 16.9 \\
120   & 4.55         & {2.67}     & 4.78       & 2.41       & 16.4 \\                             
110   & 4.14         & 2.74       & 4.37       & 2.08       & 15.8 \\
100   & 3.76         & 3.00       & 4.03       & 1.76       & 15.1 \\
90    & 3.47         & 3.63       & {3.87}     & 1.50       & 14.3 \\
80    & {3.42}       & 4.99       & 4.13       & {1.40}     & 13.4 \\                
70    & 4.02         & 7.94       & 5.41       & 1.74       & {12.4} \\
\tableline
\end{tabular}
\tablenotetext{(a)} {core-mantle-particles with a SiC core volume fraction of 0.5}
\end{center}
\end{table*}

{Despite the good spectroscopic fits obtainable with TiC, cold 
SiC+SiO$_2$ and cold FeO, we come to the conclusion, based on the properties 
of the sources of the `21'\,$\mu$m band, that none of the hitherto discussed candidate 
carriers is a convincing feature carrier candidate in all relevant respects.
Even though there are oxides which may appear as likely band carriers from
a spectroscopical point of view, it is difficult to conceive of a 
scenario accounting for their formation in carbon-rich PPNe. 
On the other hand,} dust species which are usually predicted to form in carbon-rich
environments, like carbon, carbides and sulfides do either not match the observed
`21'\,$\mu$m band profile (e.\ g.\ SiS$_2$), or even their maximum abundance 
is too small for emitting the energy contained -- at least in some cases -- 
in the `21'\,$\mu$m feature (which rules out TiC). 
{As for the individual groups of feature carriers, our arguments are the
following ones:}

1) Dust and macromolecular species which produce a 20--22\,$\mu$m emission
by transitions involving heteroatoms in carbonaceous networks (e.g.\ Papoular 2000),
are not likely to be the feature carrier due to the uniqueness of the 
`21'\,$\mu$m band profile on the one hand and the non-uniqueness of the 
heteroatom-caused emission bands on the other hand.

2) There do exist oxide dust species which can produce a 20.1\,$\mu$m emission
band very effectively (if the question of their formation in C-rich environments
is left aside). As we have shown, this holds true both for cold FeO and for
SiO$_2$. Iron monoxide has {\em only}\/ one IR active mode, centered at
19.9\,$\mu$m, which shows an exceptional {narrowing}\/ at temperatures in
the 200--100\,K range. By contrast, amorphous SiO$_2$ has two strong IR active
modes, one of them being centered at 20.5\,$\mu$m. Our laboratory experiments
show that amorphous SiO$_2$ easily forms mantles on top of SiC. Core-mantle particles
composed of SiC and SiO$_2$, if colder than 150\,K, emit mainly through the lower energy
band of SiO$_2$. Thus they can in principle account for the `21'\,$\mu$m
band, even though the reproduction of the feature profile is not as good 
as in the case of cold FeO.

But there is a major problem with any scenario involving oxides as band carriers:
the difficulty to explain how oxides can form or at least survive in carbon-rich
\mbox{PPNe}. It is hardly conceivable that the carrier of the `21'\,$\mu$m
band is just a remnant of the oxygen-rich outflow phase of the respective former 
AGB stars. The interval between the oxygen-rich phase and the carbon-rich PPN
phase is too long as to account for a co-location, as proven for HD 56126,
of products of the C-rich and the O-rich phase. The remnants of the O-rich phase
are expected to be located at least one order of magnitude further out on a radial 
scale than the products of the C-rich phase.
How, then, can either SiC be oxidized or FeO form/survive in C-rich environments?
As for the first process, there are several studies showing that {\em complete}\/ 
transformation of SiC to SiO$_2$ occurs at temperatures above 700\,K (e.g.\ Mendybaev 
et al.\ 2002, Tang \& Bando 2003); provided that large defective surfaces as those 
of nanograins are available for the reaction, the formation of SiO$_2$ mantles on top of 
SiC grains also takes place at room temperature as our laboratory experiments 
indicate (see Fig.\ \ref{TEM}). But why should this SiO$_2$ mantle formation
take place preferentially in PPNe with C/O ratios close to or larger than unity? 
This question, unfortunately, has to remain open here. 
As for the hypothesis that FeO is the carrier of the `21'\,$\mu$m band, 
it would require the oxidation of Fe to FeO in cold circumstellar
environments. It seems that metal-poor PPNe like HD 56126 -- paradoxically,
at the first glance -- provide good conditions for FeO formation, since
this process requires the presence of small metallic iron grains.
The {\em absence}\/ of any feature attributable to FeO in the vast majority 
of the Planetary Nebulae, on the other hand, could be accounted for by the 
fact that FeO can persist only in a narrow range of equilibrium between oxidation 
(to higher iron oxides) and reduction (to metallic iron). Thus, in PNe, FeO is 
probably reduced to metallic iron.

3) Quantum size effects critically influence the optical properties of very small grains and
clusters. Phonon modes can shift when the lattice changes from `infinite' to finite size, and 
conductivity may be reduced. For the TiC {\em grainlets}\/ proposed as carriers, this
means that a `21'\,$\mu$m band would only occur for clusters and particles in a 
{\em very narrow size range}\/. This makes it unlikely that there would be a sufficient 
number of TiC particles in the \mbox{PPNe} to account for the observed band strength, given 
that the predicted Ti abundance is not large enough and that mantling by amorphous 
carbon will reduce the band strength even further.

4) Any identification or ruling-out of a candidate carrier is strengthened
if it does not only have one, {\em but two or more observable infrared bands}\/. 
If these bands are seperated by $\ge$10\,$\mu$m, the secondary bands may be 
suppressed by temperature effects. This is likely to occur for the 11\,$\mu$m SiC 
band expected to arise from the core-mantle particles mentioned above. However, 
in the case of SiS$_2$, even temperatures in the order of {100\,K} will not be 
sufficient to suppress the 16.8\,$\mu$m secondary emission feature. The multiband 
identification problem also sets constraints on bands produced by impurities, 
because the impurity-induced bands are usually much weaker than the phonon 
bands of the respective matrix materials (which provides a further argument 
against doped SiC as the carrier of the '21'\,$\mu$m feature).


{\acknowledgments
We thank Gabriele Born, Jena, for preparing our samples.
Ingrid Hodou\v s, Vienna, kindly provided us with the ISO-spectra
of IRAS 07134+1005 and SAO 34504, reduced with the ISO spectral analysis 
package ISAP. An anonymous referee contributed some substantial 
arguments to the original version of this paper and thereby 
improved it. TP received a DOC grant from the Austrian Academy 
of Sciences. HM acknowledges support by DFG grant 
Mu \mbox{1164/5.}}


\appendix

\section{Appendix: On the temperature profile and spectral energy distribution
of geometrically thin PPNe dust shells}

{In the previous sections, calculations of the radiative equilibrium
temperature of the sources of the `21'\,$\mu$m feature have been presented.
The temperature profiles in turn have been partly used to calculate model spectra.
We append a discussion of the method used for the dust temperature (T$_{\rm d}$) calculation
as well as of the influence of T$_{\rm d}$ on the emerging spectral energy distributions
(SEDs).

In the radiative equilibrium case and for optically thin shells (where backwarming
effects can be neglected), the radial variation of T$_{\rm d}$ is derived from
the following energy balance:   

\begin{eqnarray}
\int_{0}^{\infty} \pi a^{2} Q_{\rm abs}(\lambda) \pi B_{\rm \lambda}(\lambda, T_{\rm eff}) \, \frac{R_{\rm eff}^{2}}{R^{2}(T_{\rm d})} \, d\lambda 
= \int_{0}^{\infty} 4 \pi a^{2} Q_{\rm abs}(\lambda) \pi B_{\rm \lambda}(\lambda, T_{\rm d}) \, d\lambda
\end{eqnarray}

The first term describes the energy absorbed by an individual dust grain with
a radius {$a$} and an absorption efficiency factor $Q_{\rm abs}(\lambda)$ in a
stellar radiation field $B_{\rm \lambda}(\lambda, T_{\rm eff})$ ($T_{\rm{eff}}$ denoting
the effective stellar temperature); the second term describes the energy 
emitted by the same dust grain. $R_{\rm eff}$ is the effective stellar radius,
whereas $R(T_{\rm d})$ is the distance of the grain with temperature T$_{\rm d}$
from the star. While the above integral runs from zero to infinity,
numerical integration can of course be performed only between a certain lower 
($\lambda_{\rm min}$) and a certain upper limit ($\lambda_{\rm max}$). We chose
$\lambda_{\rm min}$ = 0.05\,$\mu$m and $\lambda_{\rm max}$ = 500\,$\mu$m.
It is difficult, however, to derive the values of $Q_{\rm abs}$ close to
$\lambda_{\rm min}$; some of the sets of optical constants have to be 
extrapolated to UV wavelengths since measurements of the UV properties of
cosmic dust analogues are still lacking. For FeO, e.g., our measurements
extend to 0.2\,$\mu$m; beyond this wavelength, n and k were assumed to
be constant. Any other assumption, and even the choice $\lambda_{\rm min}$ 
= 0.2\,$\mu$m, would not significantly change our results. 

We recall that simple power-law models for Q$_{\rm abs}(\lambda)$
make it possible to solve eq.\ (A1) analytically for T(r). For Q$_{\rm abs}(\lambda)$
$\propto$ 1/$\lambda$ -- the case corresponding to small dust particles with
wavelength-independent (practically: weakly wavelength-dependent)
optical constants $n$ and $k$ --, T(r) decreases with 1/r$^{0.4}$. This
solution to eq.\ (A1) has been plotted in Fig.\ \ref{FeOTR}.
(For the less realistic case Q$_{\rm abs}(\lambda)$ = const., T(r) 
decreases with 1/r$^{0.5}$.)

In the main part of our paper, it has been shown that the temperature
range corresponding to the measured spatial extension of a typical PPN like
HD 56126 is comparatively narrow for all the dust species examined. The
ratio between the temperature at the outer and the temperature at the inner 
dust shell edge (i.e., its dense observable part) is typically in the order 
of 1.3. This has an important consequence for the calculation of the dust shell's
SED. Again in the optically thin case, the latter is given by

\begin{equation}
F_{\rm d}(\lambda) = c \times Q_{\rm abs}(\lambda) \int_{r=1}^{r=R_{out}/R_{in}} B_{\rm \lambda}(\lambda, T_{\rm d}(r))\,{}dr ,\\ 
\end{equation}

where $c$ is a product containing the dust column density, the
average geometrical cross section of the dust grains, and the surface of the
inner dust shell boundary (e.g.\ \citet{ST89}). If R$_{out}$/R$_{in}$ is indeed in
the order of 2 (as for HD 56126 according to the observations of \citet{KVH02} 
and \citet{Miy03}), the result of the integral on the right-hand side of eq.\ (A2)
is surprisingly similar to a single black-body function for a temperature
intermediate between T$_{\rm d}$(R$_{in}$) and T$_{\rm d}$(R$_{out}$).
This effect is shown in Fig.\ \ref{f:BB}, where we have set c=1 and Q$_{\rm abs}$($\lambda$)=1 
for simplicity.

\begin{small}
\begin{figure}[htbp]
\plotone{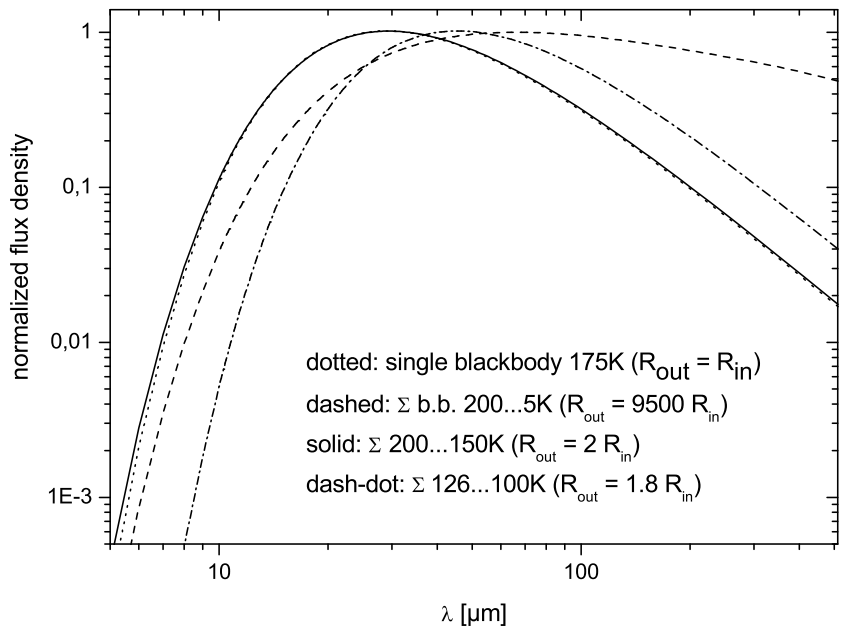}
\caption{Spectral energy distributions (F$_{\nu}$($\lambda$)) emerging from
a single blackbody (dotted line, for T=175\,K) compared to SEDs emerging from
a sum of blackbody contributions as the one entering into eq.\ (A2).
For R$_{out}$/R$_{in}$ $\le$ 2 (solid and dash-dotted line), the blackbody-sum 
is virtually not distinguishable from a single blackbody with a temperature 
intermediate beween T(R$_{in}$) and T(R$_{out}$). Only for large values of
R$_{out}$/R$_{in}$ (e.g.\ values close to 10000, as for the dashed line),
the blackbody-sum has a profile distinctly different from that of a
single Planck function.
\label{f:BB}}
\end{figure}
\end{small}

It is evident from this figure that the the SED calculated for a geometrically 
thin circumstellar shell (using realistic absorption efficiencies) according 
to eq.\ (A2) depends much more on the order of magnitude of the dust temperature 
(i.\ e.\ on whether the mean dust temperature amounts to 175 or 125\,K) than on 
actually calculating a weighted average of Planck functions.

On the other hand, even a reasonably good estimation of the mean dust temperature
T$_{\rm d}(r)$ is difficult without any information on the values of Q$_{\rm abs}$($\lambda$)
in the visual and near-infrared range. For SiS$_2$ and TiC, e.g., no optical constants
for wavelengths smaller than 10\,$\mu$m are available. Hence, the SEDs emerging from
dust shells composed of these grain species can only be calculated by arbitrarily
assuming mean T$_{\rm d}$-values for them.}



\begin{thebibliography}

\bibitem [Buss et al.(1990)] {Buss90}
   Buss, R.H., Cohen, M., Tielens, A.G.G.M., et al., 1990,
   \apj 365, L23

\bibitem [Chigai et al.(2003)] {Cetal03}
   Chigai, T., Yamamoto, T., Kaito, C., \& Kimura, Y., 2003,
   \apj 587, 771

\bibitem [Cl\'e\-ment et al.(2003)] {Clem03}
   Cl\'ement, D., Mutschke, H., Klein, R., \& Henning, Th., 2003, 
   \apj 594, 642

\bibitem [Cl\'e\-ment et al.(2004)] {Clem04}
   Cl\'ement, D., Mutschke, H., Klein, R., J\"ager, C., Dorschner, J.,
   Sturm, E., \& Henning, Th., 2004
   \apj, submitted

\bibitem [Cox(1990)] {Cox90}
   Cox, P., 1990, \aap 236, L29

\bibitem [Dayal et al.(1998)] {Day98}
   Dayal, A., Hoffmann, W.F., Bieging, J.H., Hora, J.L., Deutsch, L.K.,
   \& Fazio, G.G., 1998, \apj 492, 603

\bibitem [Duley(1980)] {Dul80}
   Duley, W.W., 1980, \apj 240, 950

{\bibitem [Fehlner \& Mott(1970)] {FM70}
   Fehlner, F.P., \& Mott N.F., 1970, Oxid. Met. 2, 59}

{\bibitem [Ferrarotti \& Gail(2003)] {FG03}
   Ferrarotti, A.S., \& Gail H.-P., 2003, \aap 398, 1029}

{\bibitem [Gail \& Sedlmayr(1998)] {GS98}
   Gail H.-P., Sedlmayr E., 1998, in: Chemistry and Physics of Molecules and
   Grains in Space, Faraday Discussions no.\ 109, The Faraday Division of the
   Royal Society of Chemistry, London, p.\ 303}

\bibitem [Goebel(1993)] {Goe93}
   Goebel, J.H., 1993, \aap 278, 226

\bibitem [Grish\-ko et al.(2001)] {Gris01}
   Grishko, V.I., Tereszchuk, K., Duley, W.W., \& Bernath, P., 2001,
   \apj 558, L129

\bibitem [Henning et al.(1995)] {H95}
   Henning, Th., Begemann, B., Mutschke, H., \& Dorschner, J., 1995
   \aaps 112, 143

\bibitem [Henning \& Mutschke(1997)] {HM97}
   Henning, Th. \& Mutschke, H., 1997, \aap 327, 743 

\bibitem [Henning \& Mutschke(2001)] {HM01}
   Henning, Th., \& Mutschke, H., 2001, Spectrochim.\ Acta 57, 815

\bibitem [Hill, Jones \& d'Hen\-de\-court (1998)] {Hill98}
   Hill, A.G.M., Jones, A. P., \& d'Hen\-de\-court, L.B., 1998,
   \aap 336, L 41

\bibitem [Hony, Waters \& Tielens(2001)] {HWT01}
   Hony, S., Waters, L.B.F.M., \& Tielens, A.G.G.M., 2001, \aap 378, L41
   
\bibitem [Hony et al.(2003)] {Hony2003}
   Hony, S., Tielens, A.G.G.M., Waters, L.B.F.M., \& 
   de Koter, A. 2003, \aap 402, 211

\bibitem [Hrivnak, Volk \& Kwok(2000)] {HVK00}
   Hrivnak, B.J., Volk, K., \& Kwok, S., 2000, \apj 535, 275

\bibitem [Justtanont et al.(1996)] {Just96}
   Justtanont, K., Barlow, M.J., Skinner, C.J., Roche, P.F., Aitken, D.K.,
   \& Smith, C.H., 1996, \aap 309, 612

\bibitem [Kimura \& Kaito(2003)] {KK03}
   Kimura, Y. \& Kaito, C., 2003, \mnras 343, 385

\bibitem [Kwok, Volk \& Hrivnak(1989)] {KVH89}
   Kwok, S., Volk, K., \& Hrivnak, B.\ J., 1989, \apj 345, L51

\bibitem [Kwok, Volk \& Hrivnak(2002)] {KVH02}
   Kwok, S., Volk, K., \& Hrivnak, B.\ J., 2002, \apj 573, 720

\bibitem [Lehmann, Schumann \& H\"ubner(1984)] {LSH84}
   Lehmann, A., Schumann, L., \& H\"ubner, K., 1984, 
   phys.\ stat.\ sol.\ B 121, 505

\bibitem[Li \& Draine(2002)] {LD02}
   Li, A., \& Draine, B., 2002, \apj 564, 803
   
\bibitem [Li(2003)] {Li03}
   Li, A., 2003, ApJ 599, L45

\bibitem [Mendybaev et al.(2002)] {Mend02}
   Mendybaev, R.A., Beckett, J.R., Grossman, L., Stolper, E., Cooper, R.F.,
   \& Bradley, J.P., 2002, \gca 66, 661

\bibitem [Meixner et al.(2003)] {Meix03}
  Meixner, M, Zalucha, A. Fong, D., \& Justtanont, K., 2003,
  in: Astrophysics of Dust, ed.\ A.N.\ Witt, poster P73

\bibitem [Miyata et al.(2003)] {Miy03}
   Miyata, T., Kataza, H., Okamoto, K.Y., et al., 2003,
   in: Astrophysics of Dust, ed.\ A.N.\ Witt, poster P74

\bibitem [Mutschke et al.(1999)] {M99}
   Mutschke, H., Andersen, A.\ C., Cl\'ement, D., Henning, Th., \& Peiter, G. 1999,
   \aap 345, 187

\bibitem [Ossenkopf, Henning \& Mathis(1992)] {Osk92}
   Ossenkopf, V., Henning, Th., \& Mathis, J.S., 1992,
   \aap 261, 567

{\bibitem [Oudmaijer \& de Winter(1995)] {OuW95}
   Oudmaijer, R.D., \& de Winter, D., 1995, \aap 
   295, L43}

\bibitem [Palik(1985-98)] {Palik85}
   Palik, E.\ D.\ (ed.) 1985-1998, Handbook of Optical Constants
   of Solids, 3 vols. (Boston MA: Academic Press)

\bibitem [Papoular(2000)] {Pap00}
   Papoular, R., 2000, \aap 362, L9

\bibitem [Posch et al.(1999)] {P99}
   Posch, Th., Kerschbaum, F., Mutschke, H., Fabian, D., Dorschner, J., 
   \& Hron, J., 1999, \aap 352, 609

\bibitem [Posch et al.(2002)] {P02}
   Posch, Th., Kerschbaum, F., Mutschke, H., Dorschner, J., \& J\"ager, C. 2002, 
   \aap 393, L7

\bibitem [Purcell(1969)] {P69}
 Purcell, E.\ M.\ 1969, \apj, 158, 433 

\bibitem [Roberts(1961)] {WR61}
   Roberts W.M., 1961, Trans.\ Faraday Soc.\ 57, 99

{\bibitem [Schutte \& Tielens(1989)] {ST89}
   Schutte, W.A., \& Tielens, A.G.G.M., \apj 343, 369}

\bibitem [Smith \& Witt(2002)] {SW02}
   Smith, T.L., \& Witt, A.N., 2002, \apj 565, 304

\bibitem [Speck \& Hof\-meister(2004)] {SH04}
   Speck, A.K., \& Hofmeister, A.N., 2004, \apj 600, 986

\bibitem [Sourisseau, Coddens \& Papoular(1992)] {Sour92}
   Sourisseau, C., Coddens, G., \& Papoular, R., 1992, \aap 254, L1

\bibitem [Suttrop et al.(1992)] {Sutt92}
   Suttrop, W., Pensi, G., \& Choyke, W.J., Stein, R.,
   \& Leibenzeder, S., 1992, J.\ Appl.\ Phys.\ 72, 3708

\bibitem [Tang \& Bando(2003)] {Tang03}
   Tang, Ch., \& Bando, Y., 2003, Appl.\ Phys.\ Lett.\ 83, 659

\bibitem [Volk, Kwok \& Hrivnak(1999)] {VKH99}
   Volk, K., Kwok, S., \& Hrivnak, B.J., 1999, \apj 516, L99

\bibitem [Volk, Xiong \& Kwok(2000)] {VXK00}
   Volk, K., Xiong, G.-Z., \& Kwok, S., 2000, \apj 530, 408

\bibitem [van Winckel \& Reyniers(2000)] {VW00}
   Van Winckel, H., \& Reyniers, M., 2000, \aap 354, 135

\bibitem [von Helden et al.(2000)] {Helden2000}
   Von Helden, G., Tielens, A., van Heijnsbergen, D., et al. 2000,
   Science 288, 313

\bibitem [Webster(1995)] {Web95}
   Webster, A., 1995, \mnras 277, 1555

\end{thebibliography}
\end{document}